# Successive Magnetic Transitions of the Kagomé Staircase Compound $Co_3V_2O_8$ Studied in Various Magnetic Fields

Yukio Yasui, Yusuke Kobayashi, Minoru Soda, Taketo Moyoshi and Masatoshi Sato*

*Department of Physics, Nagoya University, Furo-cho, Chikusa-ku, Nagoya 464-8602*

Naoki Igawa[1] and Kazuhisa Kakurai[1]

[1]*Quantum Beam Science Directorate, Japan Atomic Energy Agency, Tokai-mura, Naka-gun, Ibaraki, 319-1195*

**Abstract**

For the spin-3/2 kagomé staircase system $Co_3V_2O_8$, magnetic field ($H$)-temperature ($T$) phase diagrams have been constructed for the fields along three principal directions up to 5 T, using results of various macroscopic measurements on single crystal samples and also using neutron diffraction data taken on both powder and single crystal samples under $H$ along $c$. In zero magnetic field, the system exhibits three transitions at temperatures $T_{c1} \sim 11.2$ K, $T_{c2} \sim 8.8$ K and $T_{c3} \sim$ (6.0-7.0) K. The single crystal data present clear evidence for the noncollinear nature of the magnetic structures in all magnetically ordered phases below $T_{c1}$. The sinusoidal nature of the incommensurate modulation of the ordered moment reported in the former work has been confirmed between $T_{c1}$ and $T_{c2}$, that is, no higher harmonics of the modulation have been detected even for the present large single crystal. Even in the phase of commensurate modulation between $T_{c2}$ and $T_{c3}$, we have not detected any higher harmonics of the modulation. The phase diagrams show that the magnetically ordered phases sensitively change to other phases with $H$, indicating that the geometrical frustration inherent in this system is important for the determination of the phase diagram. No evidence for ferroelectric transitions has been observed in the measurements of the dielectric constant ($\varepsilon$) applying the electric fields along three crystallographic axes, *a, b* and *c*. Only small dielectric anomalies closely connected with the magnetic phase transitions have been found.

*Corresponding author: M. Sato: e43247a@nucc.cc.nagoya-u.ac.jp

## §1. Introduction

Spins on triangle and kagomé lattices often present good examples of geometrically frustrated systems. Theoretically, classical Heisenberg spin systems with antiferromagnetic (AF) nearest neighbor interaction on a 2D triangle lattice is expected to have the so-called 120º spin state.[1] On the other hand, such an arrangement is not expected for the kagomé lattice, because there remains too large degeneracy.[2,3] In real materials, the actual ground state may be different from that of ideal systems because of the existence of a magnetic single-ion anisotropy, lattice distortion and complicated long-range magnetic interactions, for example. Even for such materials, the appearance of the interesting magnetic behavior is expected at low temperature.

The kagomé staircase compounds $A_3V_2O_8$ (A=Co, Ni and Zn) are a kind of variations of the kagomé lattice system. In these systems, the edge sharing $AO_6$ octahedra form a staircase kagomé structure and these kagomé staircase lattices are separated by the nonmagnetic $VO_4$ tetrahedra. In Fig. 1, the schematic arrangement of $A^{2+}$ ions is shown for A=Co, for example (space group:*Cmca*[4]). In this types of structure, two crystallographycally inequivalent sites exist: The cross-tie sites, A(1) form the apices of the isosceles triangles and the spine sites A(2) form the bases of the triangles. The system consists of the A(2)-A(1)-A(2) isosceles triangles and for A=Co, the ratio of the A(2)-A(2) and A(1)-A(2) bond distances is ~1.003.[4]

For these systems, it is interesting to investigate effects of the geometrical frustration on their magnetic behaviors at low temperature. Detailed experimental studies and discussion of magnetic behavior of $Ni_3V_2O_8$ have been reported in refs. 5-9. For $Co_3V_2O_8$, the magnetic susceptibility $\chi$ and the specific-heat $C$ of polycrystalline samples[5] and single crystals[10,11] as well as results of neutron scattering study[11] have been reported. However, the magnetic properties, especially data taken with various field directions at low temperature have not been reported.

Another interesting view point on these systems is their dielectric properties. For $Ni_3V_2O_8$, the multiferroic behavior has been reported, that is, magnetic and ferroelectric transitions take simultaneously at a critical temperature.[7,9] Theoretically, the mechanisms of these multiferroic behavior are proposed to be expected in systems with the transverse spiral order of the magnetic moments,[7,8,12-14] where the spontaneous polarization appears perpendicular to the propagation vector and the spin rotation axis.

To obtain more information on the magnetic properties and their relationship with the multiferroic nature, we have first carried out magnetic, dielectric and specific-heat measurements on single crystals of $Co_3V_2O_8$ in various magnetic fields along the three crystallographic axes up to 5 T. Second, we have carried out neutron diffraction measurements of both powder and single crystal samples down to 2 K, where nontrivial magnetic structures have been found. On the basis of these experimental studies, detailed $H$-$T$ phase diagrams for three field directions are constructed and compared with those of $Ni_3V_2O_8$. The existence of many phases in the phase diagrams can be considered to be related to the geometrical frustration inherent in this type of systems. No evidence for the occurrence of the ferroelectricity has been found in the dielectric data, which is comparatively discussed with the case of $Ni_3V_2O_8$, considering the conditions theoretically proposed for the occurrence of the multiferroic state.

## §2. Experiments

Polycrystalline samples of $Co_3V_2O_8$ were prepared by the solid reaction: CoO and $V_2O_5$ were mixed with the proper molar ratio and the mixtures were pressed into pellets and sintered at 800 °C for 24 h. The obtained samples were reground and pressed into pellets and heated at 1100 °C for 24 h in flowing $N_2$ gas. Single crystals of $Co_3V_2O_8$ were grown by a floating zone method[15]: The initial mixtures were pressed into rods and sintered at 900 °C for 24 h. By using the obtained rod, the single crystals were grown with a typical diameter ~6 mmϕ and 20 mm long. Polycrystalline samples of $Zn_3V_2O_8$ were prepared by the solid reaction: ZnO and $V_2O_5$ were mixed with the proper molar ratio and the mixtures were pressed into pellets and sintered at 725 °C for 24 h. The obtained samples were checked not to have an appreciable amount of impurity phases by the powder X-ray measurements. The axes of the crystals were determined by observing the X-ray or neutron diffraction lines.

The magnetization $M$ was measured in the magnetic field $H$ up to 5 T using a SQUID magnetometer (Quantum Design) in the temperature range from 2 K to 300 K. The specific-heat $C$ was measured by the thermal relaxation method using a Quantum Design PPMS (Quantum Design). The dielectric constant ε was measured with the frequency of 1 kHz using an ac capacitance bridge (Andeen Hagerling 2850A) for single crystals of $Co_3V_2O_8$, to which the electrodes were attached with the silver paint.

Powder neutron diffraction measurements were carried out using the high resolution powder diffractometer (HRPD) installed at JRR-3 of JAEA in Tokai. The 331 reflection of Ge monochromator was used. The horizontal collimations were open(35′)-20′-6′ and neutron wavelength was 1.8233 Å. The sample was packed in a vanadium cylinder (10 mmϕ) and the cylinder was set in an Al-can filled with exchange gas. A displex type refrigerator was used to cool down the sample. The diffraction intensities were measured in the 2θ-range from 2.5 to 165° with a step of 0.05°. Rietveld analyses were carried out by using the RIETAN-2000.[16]

Neutron measurements on a single crystal were also carried out using the triple axis spectrometer HQR(T1-1) of the thermal guide installed at JRR-3 of JAEA in Tokai, where the double axis condition was adopted. The horizontal collimations were 12′(effective)-20′-60′ and the neutron wavelength was 2.4590 Å. The 002 reflection of Pyrolytic graphite (PG) was used as the monochromator. PG filters were placed after the second collimator and after the sample to suppress the higher-order contamination. In the measurements under the zero magnetic field, the crystal was oriented with the [100] and [010] axes, in one case, and the [010] and [001] ones, in another case, in the scattering plane. The crystal was set in an Al-can filled with exchange gas, and the can was attached to the cold head of the Displex type refrigerator. In the neutron measurements under the applied magnetic field, the crystal was oriented with the [100] and [010] axes in the scattering plane. The magnetic field was applied using a superconducting magnet along the vertical direction. In the analyses of the data, the isotropic magnetic form factor for $Co^{2+}$ were used.[17]

## §3. Experimental Results and Discussion

The magnetization $M$ and the specific-heat $C$ of $Co_3V_2O_8$ were measured in various magnetic fields along $a$, $b$ and $c$. In Fig. 2(a), the $T$-dependences of the magnetization $M$ measured under the condition of zero field cooling (ZFC) are shown for various magnitudes and directions of the field. In Fig. 2(b), the specific-heats $C$ of $Co_3V_2O_8$ divided by $T$, $C/T$ are shown as a function of $T$ for zero magnetic field. The data of the shaded $T$-region should not be taken seriously, because in this $T$-region the latent heat has been observed and the measurement by the thermal relaxation method was not accurate. From the $M$-$T$ and $C/T$-$T$ curves, three phase transitions have been found in zero field and the transition temperatures are 11.2 K, 8.8 K, and 6.0-7.0 K, where arrows indicate the magnetic transition temperatures. We call these temperatures $T_{c1}$, $T_{c2}$, and $T_{c3}$, respectively. At $T_{c1}$ and $T_{c2}$, a small decrease of $M$ with decreasing $T$ and anomalies in the $C/T$-$T$ curve have been observed, indicating that the system undergoes the second order transitions to antiferromagnetic phases. At $T_{c3}$, the system undergoes the first order transition to a ferromagnetic phase with decreasing $T$, as was indicated by the existence of the latent heat. The sharp increase of $M$ with decreasing $T$ also indicates that the transition is of the first order. The thermal expansions $\Delta a/a$ and $\Delta b/b$ decreases and $\Delta c/c$ increases abruptly at ~6 K with decreasing $T$ (not shown), which also indicates that the first order transition with lattice distortion takes place at this temperature (The $\Delta a/a$-$T$ curve of $Co_3V_2O_8$ is reported by the authors' group in ref. 18). In this transition region between 6 K and 7 K, Szymczak et al. reported that two phase transitions at $T$=6.6 K and 6.1 K exist,[10] and Chen et al. reported that three transitions at 6.9 K, 6.5 K and 6.2 K exist.[11] We will come back to this discrepancy later in the discussion of the neutron results.

In Figs. 2(c) and 2(d), the $C/T$ are shown for $Co_3V_2O_8$ as functions of $T$ under various magnetic fields $H//a$, where arrows indicate the magnetic transition temperatures. The transition temperatures $T_{c1}$ and $T_{c2}$ shift with increasing $H$ and merge at $H$~0.35 T. With further increasing $H$, the first order transition observed at $H$=0 becomes continuous and the transition temperatures $T_{c1}$ and $T_{c3}$ merge at around 0.5 T. The $C/T$-$T$ data were also reported in ref. 10 for $H$ =0.35 T and 0.55 T applied along $a$, but they did not observe the anomaly at $T_{c2}$ in the $C/T$-$T$ curves.

In Figs. 3(a)-3(c), the $T$-dependence of $M$ of $Co_3V_2O_8$ measured under the ZFC condition are shown for various magnetic fields (a)$H//a$, (b)$H//b$ and (c)$H//c$. The transition temperatures determined as the points at which the anomalies are found in the $M$-$T$ and d$M$/d$T$-$T$ curves, are sensitive to $H$, as was found from the $C/T$-$T$ curves. The easy axis of the magnetization is found to be parallel to $a$. The $M$ value observed for this orientation below 10K corresponds to the saturated moment of Co$^{2+}$-spins ($s$=3/2; 3 $\mu_B$) at $H$>0.5T. For $H//c$, the $M$ value under the field of $H$=5 T is about 90% of the full moment at low temperature. The $b$-axis is the hard axis, where the values of the magnetization at low temperature is just ~30% of the full moment at $H$=5 T. The anisotropy of the magnetization of Co$_3$V$_2$O$_8$ is found to be rather large.

In Fig. 4, the magnetic contribution to $C/T$, $C_{mag}/T$ are shown against $T$ under the field of 0.75 T along $a$, where the continuous transition is observed at around 10 K. The $C_{mag}/T$ data of Co$_3$V$_2$O$_8$ are estimated by subtracting $C/T$ of the nonmagnetic compound Zn$_3$V$_2$O$_8$ shown in the same figure from the $C/T$ value of Co$_3$V$_2$O$_8$. In the inset of Fig. 4, the entropies $S$ deduced by the $T$-integration of $C_{mag}(T)/T$ are shown against $T$ for the fields of 0.75 and 1.0 T ($H//a$). In the integration, $C_{mag}(T)/T$ below 3 K was estimated by extrapolating the data from the region $T$>3 K. The values of $S$ at 35 K in these magnetic fields is ~14.5 J/mol·K, only ~42 % of $3R\ln(2s+1)$, where $R$ is the gas constant and $s$=3/2. The difference between the observed value and $3R\ln(2s+1)$ is considered to be due to the Ising-like anisotropy of Co$^{2+}$ spins. (The observed $S$ value at 35 K corresponds to ~84 % of $3R\ln(2)$ expected for Ising systems.

Powder neutron diffraction studies have been carried out at several temperatures, and the neutron powder patterns are shown at 30 K (>$T_{c1}$), 8.9 K (~$T_{c2}$) and 2.7 K (<$T_{c3}$) in Figs. 5(a)-5(c). We have carried out the Rietveld analysis on the data at 30 K, using the space group $Cmca$ and obtained essentially same results as those reported in refs. 4 and 11. The indices of the main nuclear Bragg reflections are shown in Fig. 5(a). The Rietveld analyses of the data below $T_{c1}$ have also been carried out omitting the magnetic reflection data. The space group of the system is found to remain $Cmca$ in the magnetically ordered state, but the significant change of the lattice constants has been found at $T_{c3}$, where the $T$-dependences of the lattice constants are consistent with the results of the thermal expansion. At 8.9 K, in addition to the nuclear reflection peaks, the superlattice peaks due to the magnetic ordering are observed in the low angle region shown in Fig. 5(b). At 2.7 K, the superlattice peaks observed at 8.9 K disappear and the magnetic Bragg reflections are observed at the nuclear Bragg points in the low angle region.

To study the detailed $T$- and $Q$-dependences of the magnetic reflection, we have carried out the intensity measurements in the ($hk0$) and ($0kl$) planes at 30 K (>$T_{c1}$), 9.0 K ($T_{c2}$<$T$<$T_{c1}$), 7.2 K ($T_{c3}$<$T$<$T_{c2}$) and 3.7 K (<$T_{c3}$). At 9.0K and 7.2 K, we have observed the magnetic reflections at $Q$-points ($h,k\pm\delta,0$) and ($0,k\pm\delta,l$) with even $k$ in the reciprocal space, where the measured intensities at ($h,k\pm\delta,0$) are relatively weak compared with those of the other magnetic reflections. We have not detected any higher harmonics of the modulation in the $T$-region $T_{c3}$<$T$<$T_{c1}$ even for the large single crystal. At 3.7 K, the magnetic scattering peaks observed at 8.9 K go away and magnetic reflections at $Q$-points ($h',k',0$) and ($0,k',l'$) ($h'$ and $k'$=even) corresponding to the nuclear Bragg points and at $Q$-points ($h'',k'',0$) ($h''$ and $k''$=odd) can be found, where the scattering intensities at ($h'',k'',0$) ($h''$ and $k''$=odd) are relatively weak. From this information on the single crystal data, the indices of the magnetic reflections found in the powder patterns are determined as shown in Figs. 5(b)-5(c) at 8.9 K and 2.7 K, respectively. The $hk\pm\delta 0$ ($k$=even) magnetic reflections at 8.9 K and the ($h''k''0$) ($h''$ and $k''$=odd) reflections at 2.7 K cannot be detected by the powder neutron scattering due to the weakness of their intensities.

The $T$-dependence of $\delta$ of the 02-$\delta$0 magnetic reflection is shown in Fig. 6(a). The scattering intensities of this reflection are observed only in the $T$-region of $T_{c3}$<$T$<$T_{c1}$. With decreasing $T$, $\delta$ decreases from the incommensurate (IC) value of $\delta$~0.55 and has the commensurate (C) value of $\delta$=0.5 at $T_{c2}$. With further decreasing $T$, $\delta$ begins to decrease again at 7.0 K. The $T$-dependence of the scattering intensities of the 020 reflection corresponding to the fundamental and ferromagnetic reflection is shown in Fig. 6(b). The scattering intensities of the 020 reflection abruptly increase at ~6.0 K with decreasing $T$ and hysteretic behavior of the intensity-$T$ curve has been observed, indicating that the system exhibits the first order transition to the ferromagnetic state at $T_{c3}$. These results indicates that the $T$-regions of $T_{c2}$<$T$<$T_{c1}$, $T_{c3}$<$T$<$T_{c2}$ and $T$<$T_{c3}$ correspond to the antiferromagnetic IC-, antiferromagnetic C- and ferromagnetic C- phases, respectively. Considering the behavior of the 02-$\delta$0 and 020 magnetic reflections, it seems to exist intermediate states metastable in the narrow $T$-region between 6.0 K≤$T$≤7.0 K. The transitions reported by other groups in this $T$ region seem to correspond to these metastable states.

The ordering pattern of Co$_3$V$_2$O$_8$ was proposed by Chen *et al.* using the results of the powder neutron scattering.[11] In the present neutron scattering studies on a single crystal, the $hk\pm\delta 0$ ($k$=even) and the $h''k''0$ ($h''$ and $k''$=odd) magnetic reflections have been newly found at 8.9 K and 2.7 K, respectively. The existence of these magnetic reflections indicates that the magnetic structure of Co$_3$V$_2$O$_8$ is noncollinear and cannot be fully explained by the ordering pattern proposed in ref. 11. However, the ordering patterns proposed in ref. 11 reproduce the intensities distribution of the magnetic reflection with relatively strong intensity. We have analyzed the ordering pattern modifying the one proposed in ref. 11. Figure 6(c) shows, schematically, the magnetic structure obtained at 8.9 K. The Co(2)-moments are modulated sinusoidally along $b$ with the wavevector $\delta$=0.51$b^*$ (The value $\delta$=0.5 corresponds to the period of 4$b'$ in the real space, where $b'$=$b$/2 is inter planar distance of kagomé planes.) Their moment-direction is parallel to $a$. The amplitude of the Co(2)-moment modulation is estimated comparing the magnetic scattering intensities with the nuclear ones, to be about 2.8(±0.1) $\mu_B$. The moments of Co(1) sites align along the direction which deviates from the $a$ axis towards the $c$ axis by the angle ~10 º, and their ordered moments can be expressed by the sinusoidal wave with the modulation vector $\delta$=0.51$b^*$. The amplitude of the modulation is 0.8(±0.1) $\mu_B$. This state undergoes the lock-in transition at $T_{c2}$. The magnetic structure at 2.7 K (<$T_{c3}$) is schematically shown in Fig. 6(d). The direction of Co(2)-moments align parallel to the $a$ axis (collinear) and the amplitude of the ordered moments is equal to 3.0(±0.1) $\mu_B$ expected for $Co^{2+}$-spins ($s=3/2$). The Co(1)-moments align along the direction deviating from $a$ towards $c$ by an angle ~10 ° and the ordered moment is 1.9(±0.1) $\mu_B$.

Next, neutron measurements have been carried out in the ($hk$0) plane down to 2 K on a single crystal under various magnetic fields applied along $c$. Examples of the observed (0,$k$,0)-scan profiles taken under $H$=1.2, 1.5 and 2.5 T are shown in Figs. 7(a)-7(c), respectively. At $H$=1.2 T, a magnetic reflection appears with decreasing $T$ at ~10.7 K at the incommensurate position of (2-$\delta$)$b^*$ ($\delta$=0.57). With further decreasing $T$, the value (2-$\delta$) gradually decreases and the intensity of the reflection becomes smaller and vanishes at 5.4 K. Instead, a new peak with $\delta$=2/3 appears at ~6.2 K. The intensity of this new peak begins to decrease at ~4.3 K and disappears at ~2.8 K. This decrease of the intensity seems to be transferred to the peak with $\delta$=1.0, which appears at ~4.3 K. The observation of these reflections indicate that there exist two first order transitions in the $T$ region below 6.2 K. Under the fields $H$=1.5 and 2.5 T, similar successive transitions to those fou3nd with $H$=1.2 T can be seen (see Figs. 7(b) and 7(c)).

The $T$-dependences of the magnetic scattering intensities of the 02-$\delta$0 reflections observed at $\delta$=1, 2/3 and ~0.57 taken under the magnetic fields of 1.5 and 2.5 T along $c$ are shown in Figs. 8(a) and 8(b), respectively. The insets of them show the $T$-dependences of the $\delta$-values of the observed peaks under the fields of 1.5 and 2.5 T along $c$, respectively. We can clearly see the change of the scattering intensity from the IC position ($\delta$~0.57) to the position of $\delta$=1 with the intermitting phase with $\delta$=2/3. In the field $H$=2.5 T, the transition between IC($\delta$<2/3)- and C($\delta$=2/3)-phases seems to be continuous with varying $T$. The incommensurate phase ($\delta$<2/3) disappears when the field $H$ becomes higher than 3 T. With further increasing $H$, the intermediate states with 2/3<$\delta$<1 appear and the transition from $\delta$=2/3 to $\delta$=1 becomes continuous above 3.5 T.

Using the results of the magnetization and specific-heat measurements and neutron scattering studies, we have constructed detailed $H$-$T$ phase diagrams for the field directions along $a$, $b$ and $c$ as shown in Figs. 9(a)-9(c), respectively. Solid symbols are the transition temperatures determined by the magnetization and specific-heat data. The phase boundaries plotted by the open symbols are determined by the neutron scattering studies in the magnetic field $H$(//$c$). The dotted lines connecting the cross symbols are defined as the inflexion points of $M$-$T$ curves, which may not correspond to the phase boundary lines. In the $T$-region below these inflexion points, the relatively large ferromagnetic component is induced by the applied magnetic field. The abbreviations P, ICAF, CAF and CF represent paramagnetic, incommensurate antiferromagnetic, commensurate antiferromagnetic and commensurate ferromagnetic phases, respectively. The $\delta$ values are the modulation vector estimated from the position of the observed 02-$\delta$0 reflections. In the shaded area, the position of the superlattice peaks gradually changes with varying $T$ and the transition becomes continuous. In $H$//$a$, the ICAF and CAF phases disappear for $H$>0.5 T and the CF phase extends to the higher temperature region, because the ferromagnetic structure observed in zero field is stabilized by the field. The magnetic states are less sensitive to the field along $b$. The $H$-$T$ phase diagram under the field $H$//$c$ is found to be complex, where the phases with $\delta$=1 and $\delta$=2/3 exist in the region of $H$>1 T. The complicated $H$-$T$ phase diagram of $Co_3V_2O_8$ can be considered to be due to the geometrical frustration expected in the present the staircase kagomé lattice.

The $H$-$T$ phase diagram of $Co_3V_2O_8$ is quite different from that of $Ni_3V_2O_8$ reported in ref. 6, which is considered to be due to the difference of the magnetic anisotropy, magnitudes of spins, and exchange interactions between the two systems. For $Ni_3V_2O_8$, the ordering patterns of both Ni(1) and Ni(2) moments in the low temperature IC-phase (3.9 K<$T$<6.3 K) are transverse spiral structure with the modulation vector along $a$.[6] The ordering pattern is understood by considering the competition between the nearest neighbor and next nearest neighbor interactions of the Ni(2) moments along $a$.[6,8] While, the ordered Co-moments of $Co_3V_2O_8$ in the $T$-region of $T_{c3}$<$T$<$T_{c1}$ have the sinusoidal modulation along $b$ as stated above, indicating that the interlayer interactions and single-ion anisotropy have a certain role in determining the magnetic structure. Moreover, it is necessary to take into account the antiferromagnetic interaction between the nearest neighbor Co(1)-Co(1) spins along the [101] or [-101] direction, to understand the noncollinear magnetic structures in all magnetically ordered phases.

The measurements of the dielectric constants ε parallel to $a$, $b$ and $c$ of $Co_3V_2O_8$ have been carried out under various magnetic fields along the three crystallographic axes up to 5 T. No evidence for the occurrence of the ferroelectric transition or multiferroic phase has been observed in the present system. The ε-$T$ curves indicate, however, that the dielectric anomalies associated with the magnetic phase transitions exist. In Fig. 10, the dielectric constants with the electric field parallel to $c$, $\varepsilon_c$ of $Co_3V_2O_8$ are shown, for example, against $T$ for various magnetic fields along $c$. The dielectric constants $\varepsilon_c$ abruptly decrease at $T_{c3}$ with decreasing $T$ in the $H$-region of $H$≤0.8 T. The step-like behavior of the $\varepsilon_c$-$T$ curves has also been observed at the phase boundary between the phases with $\delta$=1 and $\delta$=2/3 in the $H$-region of $H$≥2 T (see Fig. 9). The small anomalies or changes of the temperature derivative of the $\varepsilon_c$-$T$ curves have been observed at temperatures indicated by the open circles. These temperatures are just on the boundary between paramagnetic- and incommensurate antiferromagnetic phases.

Mostovoy[12] showed by the Landau expansion theory that if the system exhibits the magnetic transition to the collinear and sinusoidal phase, the ferroelectricity is not induced and only the temperature derivative of the dielectric constants exhibits discontinuity at the transition temperature. He also proposed that the multiferroic state is realized in systems with the transverse spiral structure and the direction of the spontaneous polarization is perpendicular to the propagation vector and the spin rotation axis. The behavior of the dielectric constants of $Co_3V_2O_8$ and $Ni_3V_2O_8$ is consistent with his proposals.

Katsura et al. [13] proposed that the relation between the polarization $P_{ij}$ and the spin direction vectors $S_i$ and $S_j$ on the neighboring $i$ and $j$ sites is described as $P_{ij} \propto e_{ij} \times (S_i \times S_j)$, where $e_{ij}$ is the vector connecting the $i$ and $j$ sites. If the total uniform polarization $P$ (=$\Sigma P_{ij}$) is finite along specific direction, the ferroelectric state is realized. If we apply this relation to calculate $P$ ($P_{ij}$) for the neighboring -Co(1)-Co(1)- and -Co(2)-Co(2)- spins in the IC antiferromagnetic phase shown in Fig. 6(c), $P_{ij}$ becomes always zero, because these spin pairs have collinear structure. For the neighboring -Co(1)-Co(2)-Co(1)- spin chains along $c$, the local polarization $P_{ij}$ becomes finite. However, the total polarization is completely canceled and the uniform spontaneous polarization is not induced. It has a clear contrast with the case of $Ni_3V_2O_8$ with the transverse spiral structure, for which the polarization is expected by the theories.

## §4. Conclusions

We have carried out the magnetization, specific-heat and dielectric measurements on single crystals of the spin-3/2 kagomé staircase system $Co_3V_2O_8$ in various magnetic fields along the three crystallographic axes up to 5 T. Neutron measurements on a powder sample as well as a single crystal with magnetic field along $c$ have been carried out down to 2 K. On the basis of the results of the neutron scattering study, detailed magnetic structures are proposed in zero magnetic field. The ordered moments have the modulated structure with the modulation vector (0,$\delta$,0) in the antiferromagnetic phases, and the ordering pattern is noncollinear in all the ordered phases. The transverse spiral state does not exist in $Co_3V_2O_8$, which is in a clear contrast with the case of the multiferroic $Ni_3V_2O_8$. It is consistent with the fact that no evidence for the occurrence of the ferroelectric transition has been observed in the present dielectric measurements, that is, the difference between $Co_3V_2O_8$ and $Ni_3V_3O_8$ can be understood by the existing theories.

We have constructed detailed $H$-$T$ phase diagrams for the fields parallel to the three crystallographic directions $a$, $b$ and $c$. The stable phases are very sensitive to the external magnetic field. In the magnetic field $H$ along $c$, the magnetically ordered phases with $\delta$=1 and $\delta$=2/3 have been found in the $H$-region of $H$>1 T. The existence of various phases can be considered to be due to the geometrical frustration.

**Acknowledgments**

Powder neutron scattering at JRR-3 was performed within the frame of JAEA Collaborative Research Program. Neutron scattering on a single crystal was performed within the national user's program of the Neutron Science Laboratory of the Institute for Solid State Physics (NSL-ISSP). The work is supported by Grants-in-Aid for Scientific Research from the Japan Society for the Promotion of Science (JSPJ) and by Grants-in-Aid on priority areas from the Ministry of Education, Culture, Sports, Science and Technology.

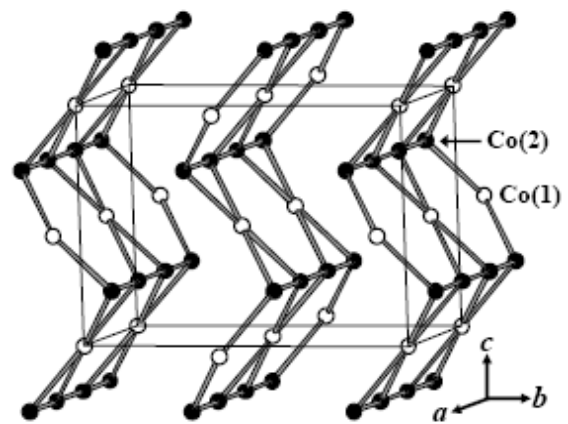

Fig. 1. Co atoms of the kagomé staircase of $Co_3V_2O_8$ are shown schematically. Open and solid circles represent the crystallographycally different Co(1) and Co(2) sites, respectively.

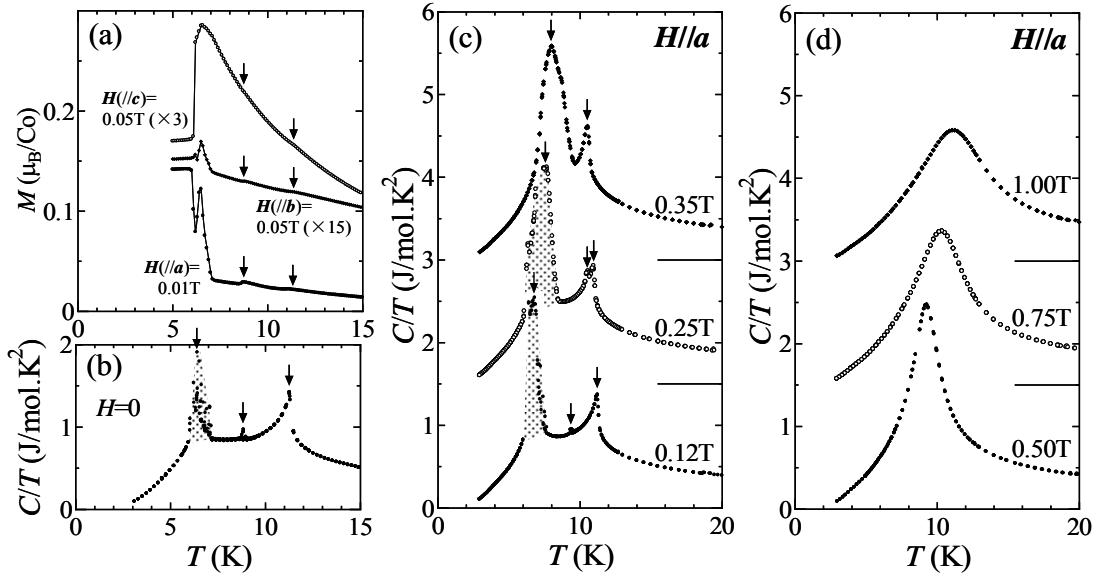

Fig. 2. (a) $T$-dependence of the magnetization $M$ of $Co_3V_2O_8$ measured under the condition of zero field cooling (ZFC) in three different field directions. Note that the values of the field are 0.01, 0.05 and 0.05 T, for the applied fields parallel to $a$, $b$ and $c$, respectively. (b)-(d) The specific-heats divided by $T$, $C/T$ are shown for $Co_3V_2O_8$ as a function of $T$ under the various magnetic field ($H//a$). Arrows indicate the magnetic transition temperatures. The data of the shaded area are taken in the temperature region of the first-order transition.

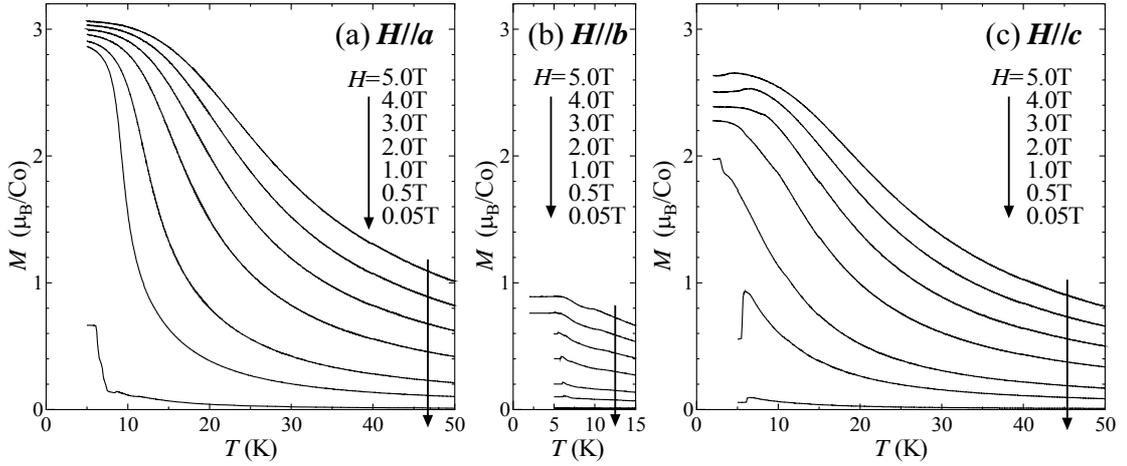

Fig. 3. $T$-dependence of the magnetization $M$ of $Co_3V_2O_8$ measured under the condition of the zero-field cooling (ZFC). The measuring fields are shown in the figures.

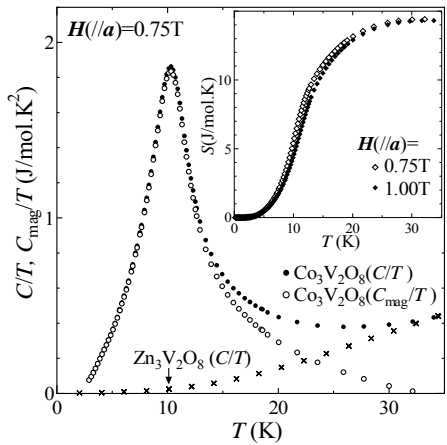

Fig. 4. The specific-heat of the magnetic contribution divided by $T$, $C_{mag}/T$ are shown for $Co_3V_2O_8$ as a function of $T$ under the magnetic field of 0.75 T along $a$. The $C_{mag}/T$ data (open circles) are obtained by subtracting the phonon contribution of the nonmagnetic compound $Zn_3V_2O_8$ (cross symbols), from the observed $C/T$ values of $Co_3V_2O_8$ (solid circles). Inset shows the $T$-dependence of the entropy $S$ deduced by the $T$-integration of $C_{mag}(T)/T$ for $H=0.75$ and $1.00$ T applied along $a$.

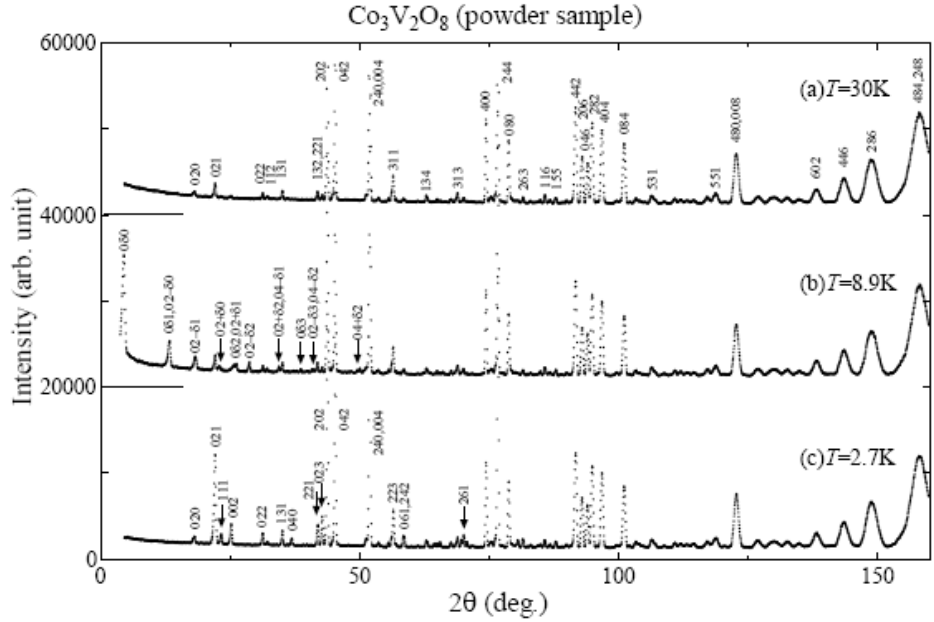

Fig. 5. Neutron powder diffraction patterns of $Co_3V_2O_8$ are shown at (a)30 K, (b)8.9 K and (c)2.7 K. The indices of the main nuclear Bragg reflections are shown at 30 K. The indices of the magnetic Bragg reflections are shown in the low angle region at 8.9 K and 2.7 K.

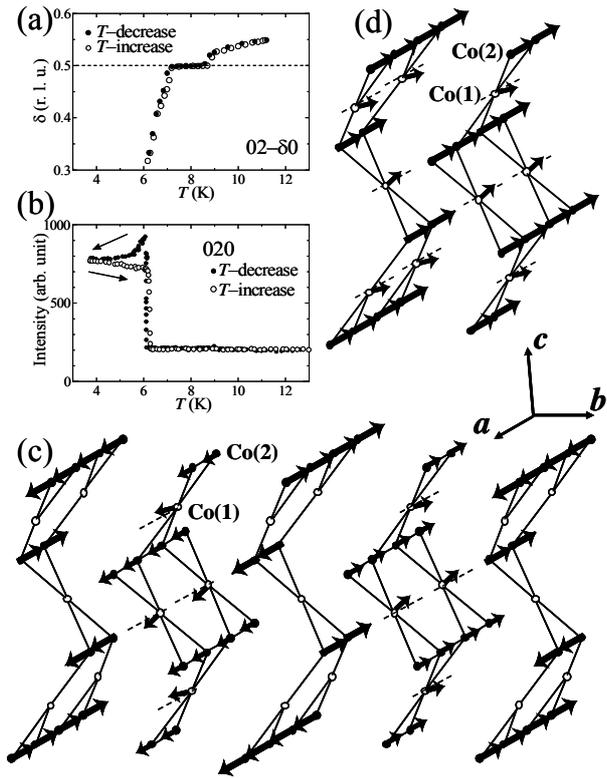

Fig. 6. (a)The $T$-dependence of the $\delta$ values is shown for $Co_3V_2O_8$. (b)The scattering intensities of the 020 reflection are shown against $T$. (c) and (d) The magnetic ordering patterns which can reproduce the observed magnetic scattering intensities of $Co_3V_2O_8$ taken at (c)$T$=8.9 K and (d)$T$=2.7 K. The thick arrows indicate the directions of the magnetic moments.

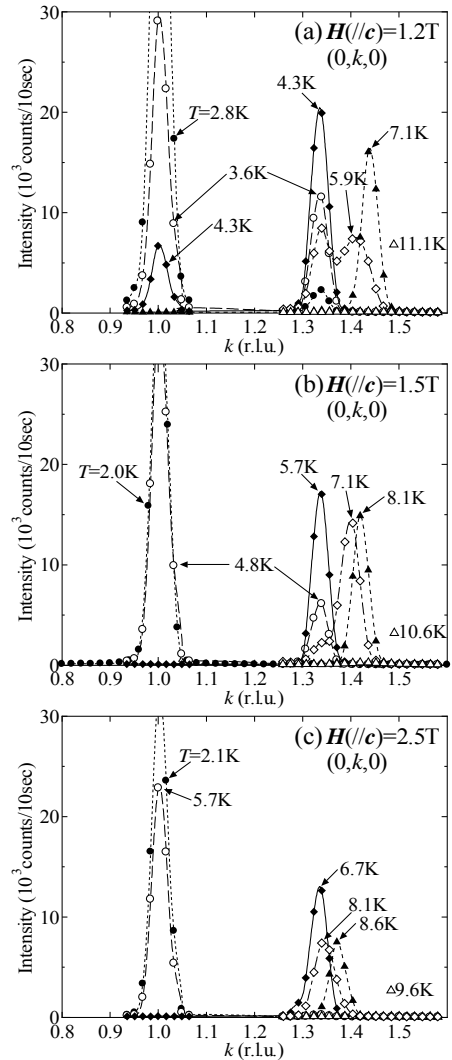

Fig. 7. Profiles of the $(0,k,0)$-scan taken under the magnetic fields (a)1.2, (b)1.5 and (c)2.5 T along $c$ are shown at various temperatures. The lines are guide for the eyes.

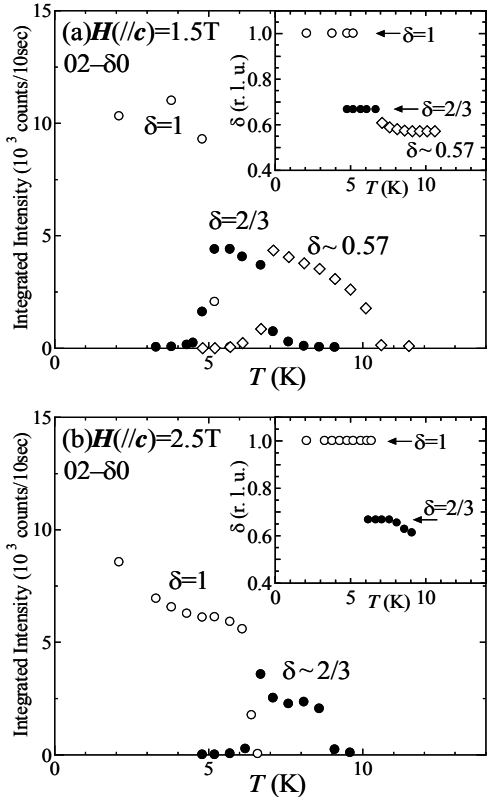

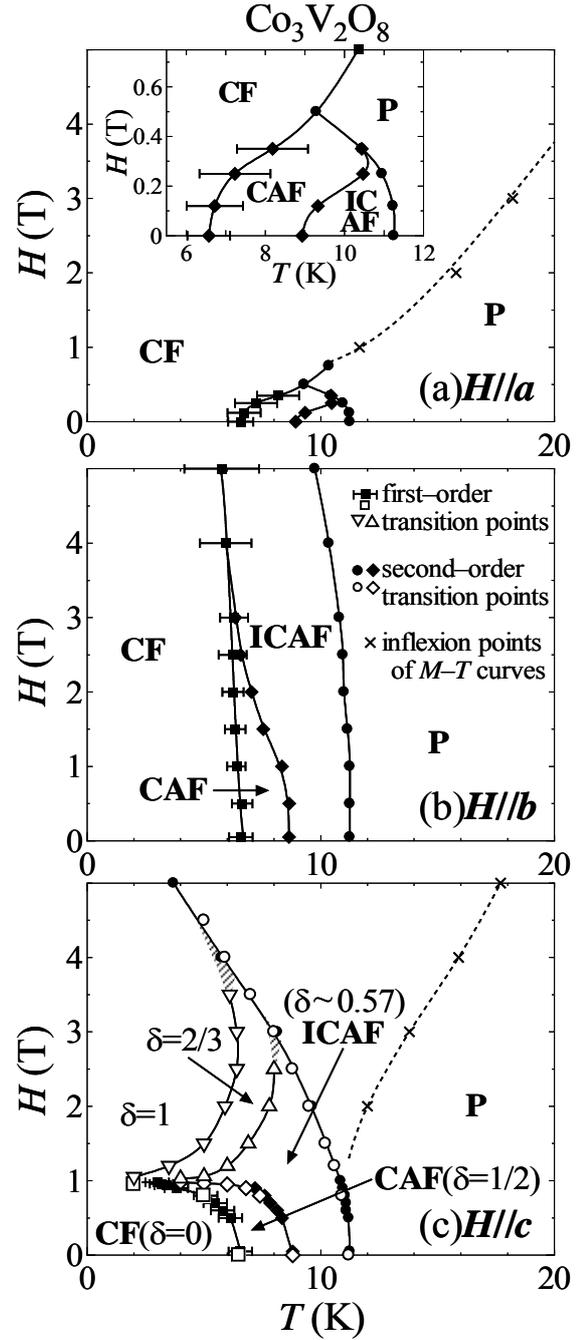

Fig. 8. The $T$-dependences of the magnetic scattering intensities of the 02-$\delta$0 reflections observed at $\delta=1$, 2/3 and ~0.57 are shown under the magnetic fields (a)1.5 and (b)2.5 T ($H//c$). The $T$-dependences of $\delta$ values of the observed peaks are shown in the inset figures.

Fig. 9. The $H$-$T$ phase diagrams of $Co_3V_2O_8$ for the magnetic fields $H$ along (a)$H//a$, (b)$H//b$ and (c)$H//c$. The abbreviations P, ICAF, CAF and CF represent paramagnetic, incommensurate-antiferromagnetic, commensurate-antiferromagnetic and commensurate- ferromagnetic phases, respectively. The $\delta$ value is estimated from the position of the 02-$\delta$0 magnetic reflections. The solid lines corresponding to the phase boundary are guides for the eyes. The dotted lines connecting the cross symbols defined as the inflexion points of the $M$-$T$ curves, may not correspond to the phase boundary lines. In the $T$-region below these inflexion points, the relatively large ferromagnetic component is induced by the applied magnetic field. In the shaded area, the position of the superlattice peak gradually changes with varying $T$ and the transition broadens and becomes continuous.

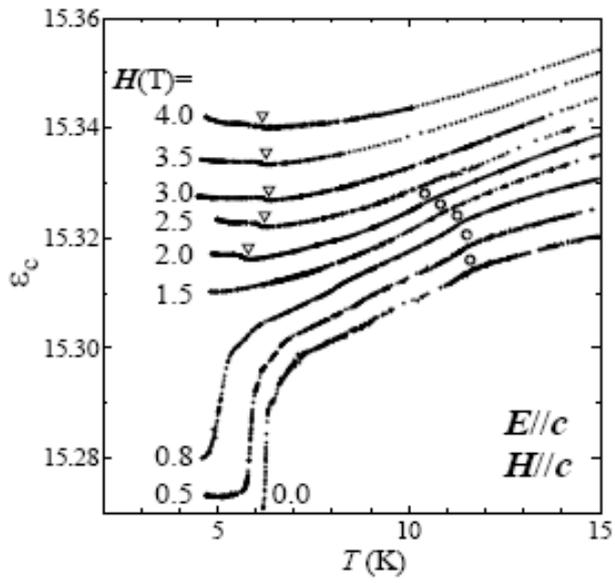

Fig. 10. The dielectric constants measured with the electric field parallel to $c$, $\varepsilon_c$ are shown for $Co_3V_2O_8$ as a function of $T$ under the various magnetic fields ($\boldsymbol{H}//\boldsymbol{c}$). The open circles indicate the boundary between P–ICAF phases. The open triangles also indicate the boundary temperatures between the phases with $\delta=1$ and $\delta=2/3$.